# Cross-Linguistic Rhythmic and Spectral Feature-Based Analysis of Nyishi and Adi: Two Under-Resourced Languages of Arunachal Pradesh

Deepshikha Gogoi[a*], Parismita Gogoi[b], Yang Saring[a]

[a] *Department of Electronics and Communication Engineering, National Institute of Technology, Arunachal Pradesh, India, 791113*

[b] *Department of Electronics and Communication Engineering, DUIET, Dibrugarh University, Assam, India, 786004*

[*]Corresponding author, National Institute of Technology, Arunachal Pradesh, India, 791113
E-mail address: gogoideepshikha16@gmail.com

# Cross-Linguistic Rhythmic and Spectral Feature-Based Analysis of Nyishi and Adi: Two Under-Resourced Languages of Arunachal Pradesh

**Abstract**

Under-resourced languages remain underrepresented in quantitative rhythm research, particularly in systematic intra-branch analysis of acoustic differentiation within closely related linguistic groups. This study investigates acoustic differentiation within the Tani language subgroup by examining speech rhythm in Nyishi and Adi, two under-resourced Tani languages spoken in Arunachal Pradesh, North-East India, using a frequency domain framework based on amplitude modulation (AM) low-frequency (LF) spectrum analysis, commonly referred to as rhythm formant analysis (RFA).The analysis is designed to identify whether intra-branch differentiation follows a hierarchical pattern across rhythmic and spectral domains. From the LF modulation spectrum, three rhythm formant features were derived: Number of Dominant peaks (NDP), Mean Frequency of Dominant Peaks (MFDP), and Variance of Dominant Frequencies (VFDP). In addition, Discrete Cosine Transform (DCT) coefficients and Mel Frequency Cepstral Coefficient (MFCC) were extracted to characterise the spectral modulation structure and broad spectral organisation of the speech signal. Statistical modelling reveals a hierarchical pattern of differentiation, where rhythmic features show consistent but moderate separation, with Nyishi exhibiting higher dominant modulation frequencies as well as greater dispersion than Adi. Classification experiments further support this hierarchy, with rhythm-only features achieved approximately 84-85% classification accuracy. Fusion using MFCC representations improved performance to 90.9% classification accuracy using support vector machine (SVM) and 93.96% using multilayer perceptron (MLP). These findings demonstrate that rhythmic and spectral features encode complementary levels of linguistic variations, with low frequency modulation capturing constrained macro temporal structure and spectral features reflecting finer phonological differentiation.

1. **Introduction**

Speech rhythm constitutes a fundamental dimension of linguistic organization, shaping temporal patterning, prominence structure, and perceptual grouping across languages. Early typological discussions differentiated languages based on stress based and syllable-based tendencies of timing; but further research has shown that rhythmic variation is more accurately characterised as a gradient and multidimensional phenomenon rather than a strict categorical division [1-2]. Linguistic research on speech rhythm spans several decades, with early accounts proposing categorical distinctions grounded in the concept of isochrony [3]. Subsequent empirical investigations questioned the acoustic validity of strict isochrony and argued for more gradient interpretations of rhythmic variation. Quantitative approaches added variability-based measures to descriptions of temporal organisation in terms of consonantal and vocalic interval distributions [4-5], which were important in empirical descriptions of rhythmic organisation. At the same time, perceptual and neuro-cognitive perspectives highlighted the importance of amplitude modulation patterns in the formation of rhythmic perception and suggest that rhythm may be coded in the acoustic signal as structured temporal modulation [6]. These developments motivate signal-based approaches that conceptualize rhythm as an emergent property of acoustic modulation.

Building on this modulation-oriented perspective, speech rhythm can be modelled as an oscillatory phenomenon reflected in the amplitude modulation (AM) structure of the acoustic signal. In this approach, rhythm appears because of reflecting the hierarchically arranged temporal fluctuations contained in the speech envelope. Frequency domain approaches therefore examine the spectral features of the low frequency (LF) modulation range, generally below 10 Hz, which show the temporal organisation at broadly syllabic and phrasal [7-8]. In the latter context, models based on rhythm formant analysis (RFA) model rhythm formant analysis based on prominent spectral peaks in the LF spectrum of the AM envelope. Rather than based on interval metric rhythm results, Rhythm Formant Analysis (RFA) is based on a final extraction of rhythmic cues tied directly to the modulating patterns inherent in the parameters of signal's amplitude envelope and avoiding the requirement of a manual segmentation of speech units [9]. This inductive methodology allows for large scale, spontaneous speech analysis and still has a sensitivity of global temporal structure.

To concretise rhythmic classification, research have traditionally attempted to search for acoustic correlates by means of the interval-based metrics. Consonant-vowel segmentation measures proposed by Ramus et al. [10] showed that a measure of segmentation variability of vocalic and consonantal intervals could be used to distinguish languages traditionally classified as stress- or syllable-timed. Subsequent refinements, including the Pairwise Variability Index [11] and related variability metrics [12], further quantified durational differences and provided evidence for rhythmic gradience rather than strict categorical typology. However, the robustness and the interpreted value of such measures have been raised. Nolan et al. [13] proposed that rhythmic dimensions might co-exist, rather than being on a straight line, and Arvaniti [14, 15] showed that metric results are very sensitive to the methodology for obtaining them and may not reflect the cross-linguistic rhythmic structure reliably. In parallel, alternative perspectives conceptualized rhythm as a dynamic and coordinated phenomenon emerging from interacting temporal processes [16,17], with additional evidence indicating that macro-prosodic and tonal patterns contribute substantially to rhythmic differentiation [18].

In response to these limitations, signal-based approaches have conceptualised rhythm as an emergent property of acoustic modulation structure to an ever-greater extent. Envelope-based analyses have shown that rhythmical organisation in languages such as English, German, Greek, Italian, Korean and Spanish could be described in terms of low-frequency modulation patterns associated with syllabic and super syllabic time scales [19]. Extending this perspective, inductive frameworks grounded in the low-frequency spectrum of the speech signal have been proposed to model prosodic organization independently of language-specific annotation, with applications to Mandarin Chinese and English [20]. Rhythm Formant Theory (RFT) and the associated methodology, Rhythm Formant Analysis (RFA), formulates rhythmic organisation at a more rigorous level using spectral peaks and trajectories in the low frequency modulation spectrum, with analyses of narrative speech in Mandarin Chinese and cross lingual comparison with English [21, 22]. Related developments such as Rhythm Zone Theory have a strong physical grounding of coexisting rhetorical rhythm zones within the amplitude envelope [23]. Recent applications have demonstrated the utility of RFA in low-resource and Indic contexts, including rhythm comparison between Mising and Assamese and classification across languages such as Bengali, Kannada, Malayalam, Marathi, Telugu, and Tamil, as well as dialect-level distinctions within Assamese [24-26]. Collectively, these studies establish modulation-based spectral modelling as a robust framework for rhythm analysis beyond traditional duration-based metrics.

The Tani languages constitute a subgroup within the Tibeto-Burman branch of the Sino-Tibetan language family and are primarily spoken in Arunachal Pradesh and adjoining regions of North-East India. This subset consists of Nyishi, Adi, Apatani, Galo and Tagin which have historical and structural affinities but exhibit unique phonological and prosodic features [27 -28]. Despite their genealogical proximity, the Tani languages show great diversity in terms of segmental inventories, tone systems and morphosyntactic units. Few annotated corpora exist for these languages and there are still few systematic analyses of their acoustic - prosodic systematics [29]. Consequently, their rhythmic organisation has received comparatively little attention in quantitative speech research especially from the modulation based or frequency domain point of view.

Within this linguistic context, the systematic study of rhythmic differentiation between closely related Tani languages has not been systematically investigated. Although previous research in North-East India has used interval-based temporal tools to study rhythm, or relied on historical comparisons across language families, modulation-based frequency domain approaches have yet to be used to discuss the intra-branch rhythmic differentiation within Tani subgroup. Languages with shared ancestry usually share prosodic and phonotactic templates, and in the Tibeto-Burman group, Tani languages such as Nyishi and Adi have predominantly CV syllable patterns and similar morphosyntactic organisation, suggesting a relative conservation of macro-rhythmic structures. Nonetheless, discrepancies in phonological realisation and segmental inventories may generate spectral divergence in the absence of restructuring in global temporal modulation patterns. From the perspective of modulation theory, one would expect relatively constrained divergence in the structure of low frequency rhythmic units (0 -5 Hz), with greater separability in spectral representations. The present study therefore investigates low frequency amplitude modulation patterns of Nyishi and Adi speech by using Rhythm Formant Analysis. By combining statistical modelling and machine-learning-based assessment, it considers whether intra-branch differentiation follows a hierarchical pattern, with more extravagant divergence of spectral organisation compared to low-frequency temporal modulation. The contribution of this study lies in providing a unified signal-based framework for intra-branch acoustic differentiation, demonstrating a hierarchical organisation across rhythmic and spectral domains in closely related under-resourced languages.

The rest of this paper is organized as follows: Section 2 presents the speech corpus and data preparation, Section 3 introduces the proposed approach including feature extraction and experimental setup, Section 4 introduces and discusses the results and Section 5 concludes this paper with comments on future studies.

## 2. Speech Corpus

The corpus was designed for ease of quantitative analysis of patterns in low-frequency amplitude modulation as well as rhythmic structure, and recordings were made in a controlled environment to ensure acoustic uniformity for subsequent frequency domain analysis. The corpus includes speech recordings of native speakers and includes spontaneous speech, reading selections and continuous narration to capture the natural as well as structured rhythmic patterns.

### 2.1 Corpus Composition: Languages and Participants

The corpus contains speech recordings of native speakers of Nyishi and Adi, two closely related languages of the Tani subgroup of the Tibeto-Burman language family. Nyishi speakers are spread largely in the districts of Papum Pare, Kra Daadi, Kurung Kumey, East Kameng, Kamle, Keyi Panyor and located in the Lower and Upper Subansiri district of Arunachal Pradesh and Adi speakers resides mainly in the district of East Siang, Siang and Upper Siang along with some in the district of Lower Dibang Valley. Although close genetically, the two languages have divergent phonological system and prosodic patterns. The corpus contains speech recordings of 20 native Nyishi speakers (12 males and 8 females) and 32 native Adi speakers (14 males and 18 females). All speakers were adult native speakers with no reported speech or hearing impairments. Table 1. lists the distribution of speakers and utterances in the Nyishi-Adi speech corpus, including gender distribution and total duration for each language. Although the number of speakers is unequal for the two languages, the corpus is balanced at the utterance level to provide equal representation for each class in the analysis. The final corpus consists of 1646 utterances of 10 seconds each, amounting to a total duration of approximately 4.57 hours of speech. Each language is represented by 823 utterances, amounting to approximately 2.29 hours of speech per language. These utterances are used as the basis for the subsequent rhythmic analysis based on modulation.

Table 1. Summary of the Nyishi-Adi Speech Corpus

| Language | No. of speakers | Male | Female | No. of Utterances | Total Duration(hours) | Avg. Utterance Length |
|---|---|---|---|---|---|---|
| Adi | 32 | 14 | 18 | 823 | 2.29 | 10 |
| Nyishi | 20 | 12 | 8 | 823 | 2.29 | 10 |
| Total | 52 | 26 | 26 | 1646 | 4.57 | 10 |

### 2.2 Data Preparation

The speech data was recorded at a sampling rate of 16 kHz from native speakers of Nyishi and Adi communities. The data was stored as WAV format. The corpus is made up of spontaneous speech, prose readings and continuous narratives aimed at reflecting both natural and structured rhythmic patterns. Manual checks were performed to remove segments where a large amount of noise or nonspeech artifacts occur. Subsequent segmentation of the audio into fixed desen united of 10 seconds produced a balanced and effectively sampled of 1,646 segments which can be accomplished for a rhythmic analysis using a low-frequency modulation technique.

### 3. Methodology

This study adopts a signal-based approach based on Rhythm Formant Analysis (RFA) to analyse speech rhythm. Moving away from traditional phonetic segmentation or interval based rhythmic analysis the present paradigm is centred on the attempted direct analysis of the spectral features of the amplitude envelope of the speech signal. By processing each utterance to extract its envelope representation and analysing it to look at the rhythmic patterns in the low - frequency band of 0-5Hz, the underlying rhythm patterns are captured. This approach reduces segmentation biases and can trace rhythm at the syllabic and phrasal level of successful segmentation. Statistical and machine learning methods are then used to judge the rhythmical differences between the Nyishi and Adi communities. Statistical and machine learning methods are then used to assess the rhythmic differences between the Nyishi and Adi communities. Figure 1. Depicts the overall framework of the Nyishi-Adi rhythm analysis system, which incorporates RFA-based feature extraction, statistical analysis, and supervised validation.

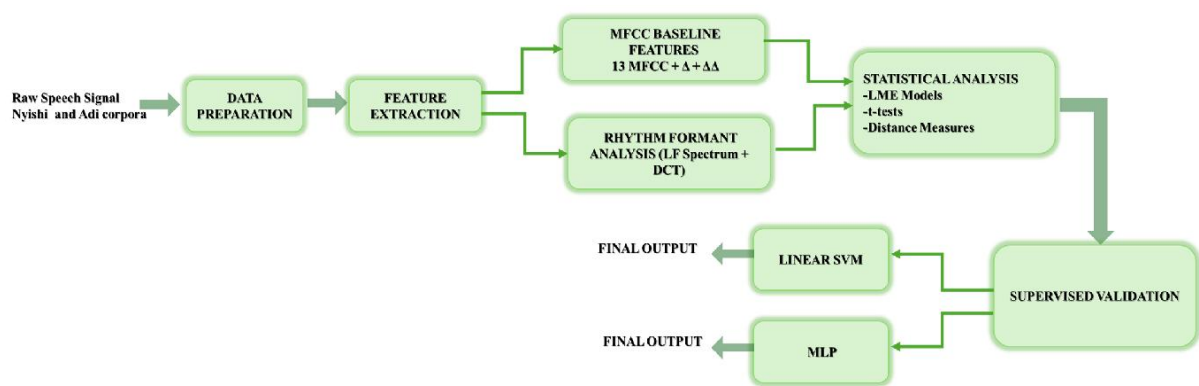

Fig.1. Computational workflow of the Nyishi-Adi rhythm analysis study

## 3.1 Rhythm Formant Analysis Framework

The current study uses the Rhythm Formant Analysis (RFA) framework for direct modeling of speech rhythm from the acoustic signal [21,23]. RFA models rhythm as patterns of low-frequency amplitude modulation and extracts rhythmic structure based on spectral prominence in the modulation spectrum. Unlike interval-based measures that require phonetic segmentation, RFA is a purely signal-level approach that eliminates segmentation bias and variability related to annotation

All analyses were performed on broadband speech recordings sampled at 16 kHz. Each utterance of 10 seconds was analysed as an independent unit. Rhythmic structure was extracted automatically from the amplitude envelope, low-frequency spectrum, and feature extraction. In the current study, only amplitude modulation (AM) features were analysed, and frequency modulation (FM) representations were not included for a specific analysis of rhythmic structure based on the amplitude envelope.

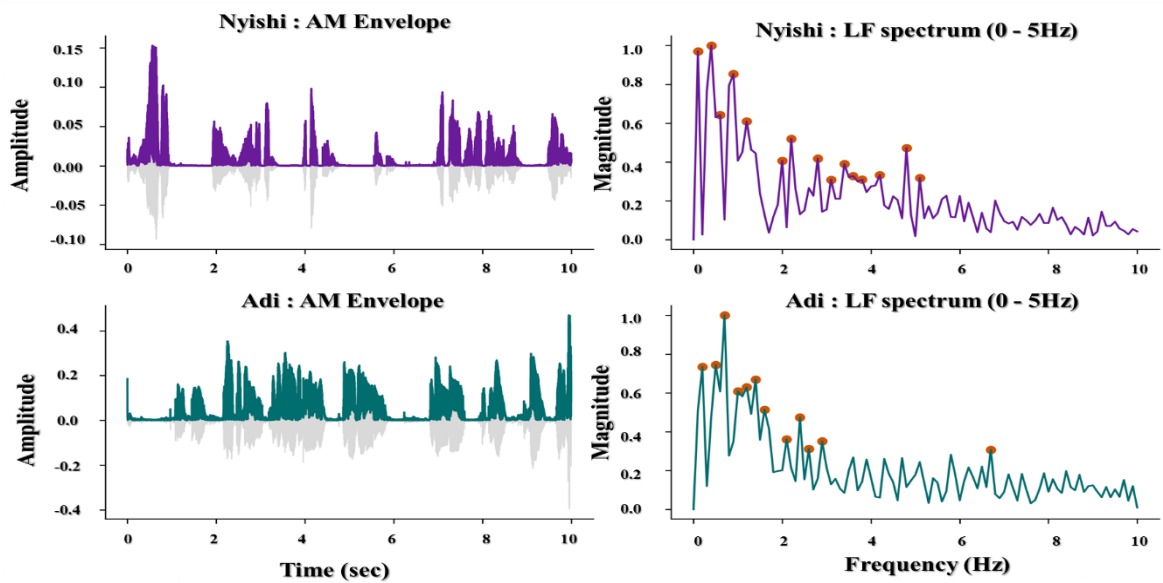

Fig.2. Amplitude envelope extraction and low-frequency (0-5 Hz) modulation spectra for representative Nyishi and Adi utterances.

## 3.2 Amplitude Envelope and Low Frequency Spectrum Extraction

For each utterance, the broadband waveform was transformed into a normalized mono signal, and peak normalization was performed to normalize amplitude scaling across utterances. The amplitude envelope was extracted by applying full-wave rectification and low-pass filtering. More specifically, the absolute value of the waveform was filtered using a 5th-order

Butterworth low-pass filter with a cutoff frequency of 5 Hz to extract slow temporal modulations characteristic of rhythmic amplitude modulation. The envelope was then normalized to the range [0, 1] using min-max scaling.

Spectral decomposition of the normalized envelope was carried out using the real-valued Fast Fourier Transform (FFT). Although the full modulation spectrum extended to the Nyquist frequency, only the low-frequency spectrum between 0 and 5 Hz was retained for rhythmic analysis as per the RFA [22,30] configuration parameters. The low-frequency amplitude modulation spectrum thus obtained served as the input for feature extraction. An example of the amplitude envelope and low-frequency modulation spectrum for representative Nyishi and Adi utterances is provided in Figure 2. The amplitude modulation spectrum was computed within the 0-10 Hz frequency range; however, rhythm formant extraction and statistical analysis were conducted within the 0-5 Hz band reflecting the dominant syllabic modulation frequencies.

### 3.3 Rhythm Feature Extraction of Nyishi and Adi

Rhythm-related feature descriptors were extracted from the modulation spectrum, the low-frequency spectral decomposition of the amplitude envelope. The feature extraction process aimed to capture discrete spectral emphasis patterns as well as global spectral shape information in the 0-5 Hz frequency range. Two different modelling approaches were adopted: (i) peak-based feature descriptors quantifying the emphasis of modulation frequencies and their distribution, and (ii) Discrete Cosine Transform (DCT)-based coefficients modelling the global spectral shape. Additionally, Mel-Frequency Cepstral Coefficient (MFCC) feature extraction was performed to provide a spectral baseline representation for comparative analysis.

#### 3.3.1  Peak-Based Rhythm Features

Peak-based rhythm feature descriptors were extracted from the prominent low-frequency amplitude modulation components identified in the 0-5 Hz spectrum. For each utterance, the six most prominent spectral peaks were chosen based on the RFA parameter settings. To evaluate the influence of varying levels of emphasis, magnitude-based thresholding was performed over a range of threshold values (0.0-0.6). Only those peaks with normalized magnitudes greater than or equal to the threshold value were considered for feature extraction.

Based upon the retained peak frequencies [31,32], the following three descriptors were computed: (i) Number of Dominant Peaks (NDP) = the number of peaks which surpass a predefined threshold; (ii) Mean Frequency of Dominant Peaks (MFDP) = average of the retained peak frequencies; and (iii) Variance of Dominant Peak Frequencies (VFDP) = measure of the dispersion of the dominant modulation frequencies.

### 3.3.2 DCT- Based Spectral Modelling

To characterise the global structural pattern of the spectrum of low frequency amplitude modulation, the Discrete Cosine Transform (DCT) was used on the normalised spectral magnitude. The DCT [33,34] gives a compressed representation of the spectral structure by a projection of the modulation spectrum onto an ordered basis of cosine functions. While peak descriptors describe local spectra phenomena, the DCT coefficients describe the overall energy distribution of the rhythmic energy within the low frequency band.

The DCT coefficients were computed as:

$$X_k = \alpha_k = \sum_{n=0}^{N-1} x_n \cos\left(\frac{\pi}{n}\left(n + \frac{1}{2}\right)k\right), k = 0,1,\ldots,N-1 \quad \text{(i)}$$

where the normalization factor $\alpha_k$ is defined as:

$$\alpha_k = \sqrt{\frac{1}{N}}, k = 0; \sqrt{\frac{2}{N}}, k > 0 \quad \text{(ii)}$$

Here, $x_n$ represents the normalized low-frequency spectral magnitude at bin $n$, $N$ denotes the total number of spectral bins within the 0-5 Hz range, and $X_k$ is the resulting DCT coefficient. The first four coefficients (DCT0-DCT3) were retained for subsequent analysis, as they capture the dominant components of the low-frequency spectral contour.

### 3.3.3 MFCC-Based Spectral Representation

As a spectral baseline feature, Mel-Frequency Cepstral Coefficients (MFCCs) were derived from each utterance using conventional short-time processing techniques [36]. The speech signal was analysed using 25ms Hamming-windowed frames with a 10ms frame shift at a sampling rate of 16 kHz. A 26-channel Mel filter bank was applied to the magnitude spectrum, and the logarithm of filter bank energies was transformed using the Discrete Cosine Transform to derive 13 MFCCs per frame.

To capture temporal dynamics, first-order (Δ) and second order (ΔΔ) derivatives were calculated, resulting in 39 coefficients per frame. For utterance-level representation, the mean and standard deviation of each coefficient were calculated over time, resulting in a 78-dimensional feature vector per utterance. These MFCC-based features provide a spectral baseline for comparison with modulation-based rhythm features.

### 3.4 Statistical Analysis

Statistical analyses were carried out at the utterance level to compare cross-linguistic differences between Nyishi and Adi. For peak-based rhythmic features (NDP, MFDP, and VFDP), independent-samples comparisons were carried out using Welch's t-test [38] for each spectral threshold (0.1-0.6), as the test does not require equal variances between groups. Mean difference estimates (Nyishi - Adi) and p-values were calculated separately for each feature and threshold. To control for speaker-level variability, Linear Mixed-Effects (LME) [37] models were fit for all acoustic feature sets, including peak-based features, DCT coefficients (DCT0–DCT3), and MFCC-based features. For MFCC-based features, utterance-level features were first standardized and reduced using Principal Component Analysis (PCA), and LME models were then applied to the first two principal components (PC1 and PC2). Threshold analysis was carried out by iterating statistical analyses over all spectral thresholds (0.1-0.6) to assess the stability of cross-linguistic separation as a function of peak detection sensitivity.

### 3.5 Distributional and Classification Analysis

In addition to statistical testing, distributional separability between Nyishi and Adi was evaluated using Bhattacharyya Distance to quantify feature overlap Supervised classification tasks were also performed to test the separability of the derived features in a speaker-independent setting. Linear and neural networks were used to compare the performance of classification tasks using different sets of features.

### 3.5.1 Bhattacharyya Distance

To measure the extent of distributional overlap between Nyishi and Adi, Bhattacharyya Distance was calculated under a multivariate Gaussian model. For low-frequency rhythm features, the two-dimensional feature space defined by MFDP and VFDP (at threshold 0.3) was considered. For spectral modelling features, the four-dimensional space defined by DCT0-DCT3 was used. For each set of features, class-specific mean vectors and covariance matrices were estimated separately for Nyishi and Adi. A regularization constant was included in the covariance matrices to promote numerical stability in inversion and determinant calculations. The Bhattacharyya Distance was then calculated using the common covariance matrix, providing a multivariate separability measure between the two languages. Higher values of Bhattacharyya Distance measure lower distributional overlap and higher separability between the two languages.

### 3.5.2 Supervised Classification.

Supervised classification experiments were carried out using a linear Support Vector Machine (SVM) as the baseline model and a shallow Multi-Layer Perceptron (MLP) as the neural classifier [39, 40]. Classification experiments were performed using 5-fold Group K Fold cross-validation in a speaker-independent fashion. The speech samples of a particular speaker were restricted to a single fold, such that no speaker was present in both the training and test sets. Both individual feature sets (LF, DCT, and MFCC) and various fusion settings were used to explore the complementary utility of rhythm, spectral, and cepstral features. The feature vectors were normalized before the model was trained to maintain a consistent scale. The SVM model used a linear kernel, and the MLP had two hidden layers with 64 and 32 units, respectively, using ReLU activation and dropout to prevent overfitting. The MLP was trained with the Adam optimizer with early stopping on validation loss. The classification results were reported in terms of mean Accuracy and Macro-F1 scores over cross-validation folds.

### 4. Results

This section discusses the results of the acoustic comparison between Nyishi and Adi in terms of quantitative measures. The results are discussed in terms of low-frequency rhythm features, DCT-based spectral features, and MFCC features. Group level differences were analysed using threshold-based comparisons, linear mixed-effects models, measures of distributional separability, and classification accuracy. The results are organized in a way that facilitates the assessment of rhythmic similarity and acoustic multidimensional differentiation between the two languages

### 4.1 Threshold-Dependent Rhythmic Variation

Low-frequency modulation features were used to compare the rhythmic differences between Nyishi and Adi for spectral thresholds between 0.1 and 0.6. The means (± SD), Nyishi-Adi differences (Δ), and independent samples Welch's t-test results for NDP, MFDP, and VFDP at spectral thresholds 0.1 to 0.6 are shown in Table 2. Significant differences are found for MFDP and VFDP at all thresholds, and for NDP from threshold 0.2. Three peak-based features were compared: Number of Dominant Peaks (NDP), Mean Frequency of Dominant Peaks (MFDP), and Variance of Frequency of Dominant Peaks (VFDP). For all thresholds, MFDP and VFDP showed language-level separation. Boxplots of NDP, MFDP, and VFDP for Nyishi and Adi at threshold 0.3 are shown in Figure 3.

Nyishi demonstrated higher MFDP values approximately 2.54–2.57 Hz than Adi values approximately 0.78–0.97 Hz. Correspondingly, VFDP values were also higher for Nyishi approximately 1.12–1.28 Hz than for Adi approximately 0.24–0.41 Hz. Independent-samples comparisons revealed significant differences ($p < .0001$) for MFDP and VFDP at all thresholds. By contrast, NDP showed little differentiation at the lowest threshold (0.1), with significant differences emerging at threshold 0.2 and above. The difference between Nyishi and Adi became greater at a higher threshold, suggesting that threshold sensitivity is more relevant to peak-count measures than frequency-based parameters. Taken together, these findings suggest that despite similar peak counts under conservative thresholding, there are systematic differences in modulation frequency distributions between the two languages, therefore, signalling subtle but consistent rhythmic differences on the frequency level.

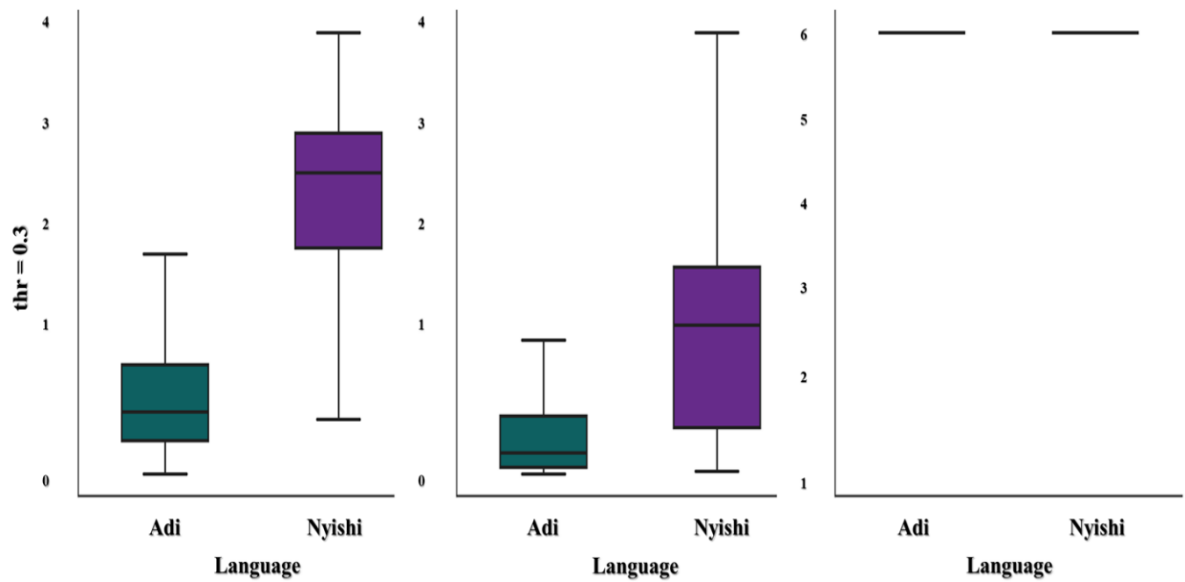

Fig.3. Distribution of rhythmic formant features at threshold 0.3

Table 2. Threshold-dependent descriptive statistics and pairwise comparisons of low-frequency rhythmic features for Nyishi and Adi.

| Threshold | Feature | Nyishi (Mean ± SD) | Adi (Mean ± SD) | Δ (Ny-Adi) | p-value |
|---|---|---|---|---|---|
| 0.1 | NDP | 6.000 ± 0.000 | 6.000 ± 0.000 | 0.0000 | n.s. |
|  | MFDP | 2.544 ± 1.041 | 0.978 ± 0.688 | 1.5664 | <0.0001 |
|  | VFDP | 1.285 ± 0.873 | 0.413 ± 0.441 | 0.8714 | <0.0001 |
| 0.2 | NDP | 5.999 ± 0.035 | 5.954 ± 0.324 | 0.0450 | <0.0001 |
|  | MFDP | 2.544 ± 1.041 | 0.971 ± 0.693 | 1.5727 | <0.0001 |
|  | VFDP | 1.285 ± 0.873 | 0.404 ± 0.437 | 0.8802 | <0.0001 |
| 0.3 | NDP | 5.983 ± 0.184 | 5.571 ± 1.047 | 0.4119 | <0.0001 |
|  | MFDP | 2.542 ± 1.045 | 0.923 ± 0.714 | 1.6190 | <0.0001 |
|  | VFDP | 1.283 ± 0.876 | 0.359 ± 0.415 | 0.9233 | <0.0001 |
| 0.4 | NDP | 5.910 ± 0.485 | 5.090 ± 1.564 | 0.8202 | <0.0001 |
|  | MFDP | 2.537 ± 1.053 | 0.871 ± 0.742 | 1.6660 | <0.0001 |
|  | VFDP | 1.279 ± 0.882 | 0.321 ± 0.410 | 0.9572 | <0.0001 |
| 0.5 | NDP | 5.657 ± 0.945 | 4.611 ± 1.900 | 1.0462 | <0.0001 |
|  | MFDP | 2.527 ± 1.090 | 0.826 ± 0.764 | 1.7003 | <0.0001 |
|  | VFDP | 1.246 ± 0.906 | 0.287 ± 0.401 | 0.9592 | <0.0001 |
| 0.6 | NDP | 5.094 ± 1.468 | 4.052 ± 2.000 | 1.0413 | <0.0001 |
|  | MFDP | 2.569 ± 1.148 | 0.783 ± 0.778 | 1.7858 | <0.0001 |
|  | VFDP | 1.127±0.948 | 0.243±0.379 | 0.8841 | <0.0001 |

### 4.2 Linear Mixed-Effects Analysis of Language Effects

Linear mixed-effects (LME) models were estimated on all acoustic feature domains, namely low frequency rhythm parameters (MFDP, VFDP, NDP), DCT coefficients of where DCT0-DCT3 and MFCC-derived principal components (PC1-PC2), with language Nyishi vs Adi was used as a fixed effect and Speaker was modelled as random intercept. As shown in Table 3, results showed that there was a significant language effect for VFDP β = 0.315, p = 0.0076, indicative of increased dispersion of dominant modulation frequencies in Nyishi than Adi, while MFDP showed a positive but marginal effect β = 0.449, p=0.060. The NDP model was not included because of near zero variance. Within the modulation domain, only DCT3 showed a significant effect β = 0.093, p = .0028, whereas while DCT0, DCT1 and DCT2 showed either no significant effects or were unstable. For MFCC-derived components, components, PC1, which explained 34.8% of total variance, showed a significant language effect β = 3.698, p = .020, while PC2 was not significant. Notably, PC1 also showed substantial speaker-level variance, reflecting pronounced inter-speaker differences in spectral structure. Altogether, the analysis extracted shows that language differentiation is best evident in VFDP, DCT3, and MFCC- PC1, reflecting the idea that rhythmic frequency dispersion and higher-order spectral organization contribute more strongly to cross-linguistic separation than simple peak-count measures.

Table 3. Fixed-Effect Estimates from Linear Mixed-Effects Models

| Feature | β (Nyishi vs. Adi) | Standard Error | p-value | Speaker Variance |
|---|---|---|---|---|
| MFDP | 0.449 | 0.239 | 0.060 | 0.691 |
| VFDP | 0.315 | 0.118 | 0.0076 | 0.146 |
| NDP | - | - | - | - |
| DCT0 | 0.155 | 0.197 | 0.433 | 0.466 |
| DCT1 | -0.238 | 0.158 | 0.133 | 0.304 |
| DCT2 | - | - | - | - |
| DCT3 | 0.093 | 0.031 | 0.0028 | 0.009 |
| MFCC-PC1 | 3.698 | 1.593 | 0.020 | 31.034 |
| MFCC-PC2 | 0.667 | 0.486 | 0.170 | 2.878 |

### 4.3 Distributional Separability Across Acoustic Feature Domains

In the rhythm domain, at a threshold level of 0.3, Nyishi speakers form a rather compact cluster centred around values of the MFDP at higher values, while Adi speakers form a more elongated distribution skewed towards lower modulation frequencies. However, there is still considerable overlap along VFDP and NDP dimensions, which results in a decent Bhattacharyya distance of 0.6457. There still is some central overlap BD = 0.7106, but the DCT feature space allows for somewhat better discrimination especially in the direction of the DCT2 dimension. In contrast, the MFCC - PCA space offers considerable global separation, mainly along the direction of PC1 with 34.84 % of total variance so that Adi extends into strongly negative values and Nyishi occupy more positive and vertically dispersed regions. PC2 with 7.85% and PC3 with 4.95% provide additional but secondary structuring. As a result, the MFCC-PC subspace is shown to achieve a significantly larger value of Bhattacharyya distance at 1.8321, which represents higher multivariate separability compared with rhythmic and modulation domains.

Overall, the findings demonstrates only a few degrees of divergence in low frequency rhythm, moderate differentiation in DCT-based modulation and dominant language's specific structuring in the whole spectral representation. Figure 4 gives a three-dimensional view of the low-frequency rhythm space MFDP, VFD, NDP , the MFCC principal component space PC1, PC2, PC3 and the DCT modulation space DCT0, DCT2, DCT 2.

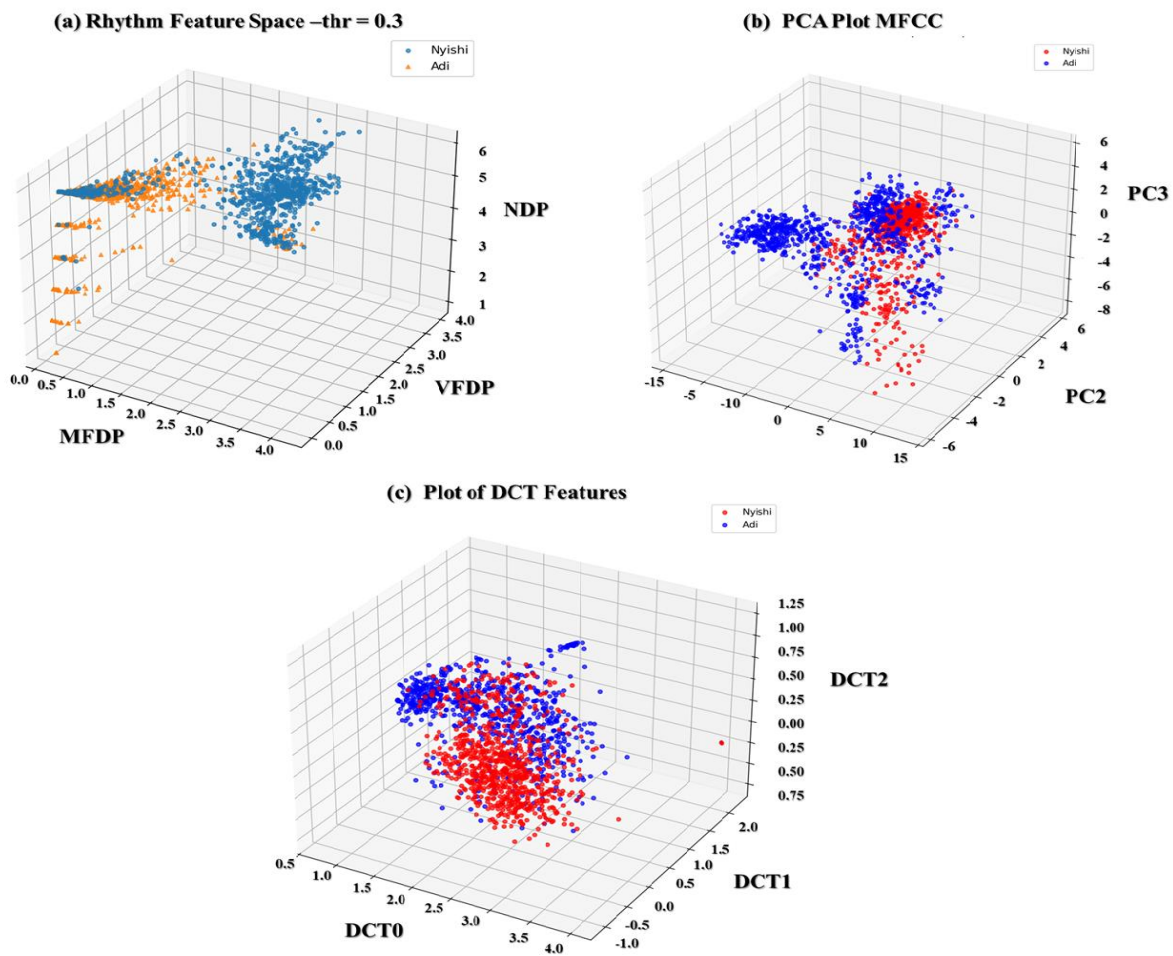

Fig.4. Multivariate Distributional Structure of Nyishi and Adi Across Rhythm, Spectral Modulation, and MFCC Feature Spaces

## 4.4 Classification Performance

To further assess language separability, supervised classification experiments were conducted using linear Support Vector Machines (SVM) and a Multilayer Perceptron (MLP). While previous analyses have described Nyishi and Adi statistically and distributionally, the more concrete description of the discriminative power of these acoustic features can be obtained through the direct measurement of classification accuracy. Accuracy was evaluated for low frequency rhythmic features, Discrete Cosine Transform modulation coefficients, Mel Frequency Cepstral Coefficient and fusion of different features.

### 4.4.1 Linear SVM Performance

Linear SVM classification performance showed clear differences in discriminative capacity across the acoustic features. Among individual feature sets, MFCC features achieved the highest performance with accuracy = 90.38% ± 3.70; macro-F1 = 87.28% ± 4.38, substantially outperforming low-frequency rhythm with accuracy 84.72% ± 7.69; macro-F1 = 62.02% ± 15.21 and DCT modulation features with 85.54% ± 6.83; macro-F1 = 63.12% ± 15.30.

Feature fusion further improved classification accuracy, with MFCC + DCT yielding the strongest accuracy = 90.90% ± 4.47; macro-F1 = 88.32% indicating the complementary role of spectral and modulation information. In contrast, combinations involving low-frequency rhythm features showed a relatively better performance. Overall, the result of linear SVM classification indicates that the spectral features, that spectral representations particularly MFCC-based features provide most robust acoustic feature sets for language discrimination. A detailed result of performance comparison of SVM and MLP classifier, for different feature combination is presented in Table 4.

### 4.4.2 MLP Performance

The MLP classifier consistently outperformed the linear SVM, especially for feature sets dominated by spectral information. Among individual domains, MFCC features again achieved the highest performance with accuracy = 91.69% ± 3.97; macro-F1 = 87.66% ± 6.25, clearly surpassing low-frequency rhythm accuracy 84.15% ± 4.45; macro-F1 = 61.88% ± 16.00 and DCT modulation features with accuracy 84.22% ± 7.03; macro-F1 = 62.45% ± 16.73. The relatively low and highly variable Macro-F1 values for LF and DCT indicate unstable class discrimination when rhythmic or modulation cues are used alone.

Feature fusion significantly improved classification accuracy, with the MFCC + DCT combination achieving the highest overall performance accuracy with 93.96% ± 4.16 and macro-F1 = 92.42% ± 4.12, which represents the strongest language discrimination achieved across all the experiments. Although with the addition of MFCC features to LF improved accuracy with 92.06%, rhythm-dominant combinations showed larger variability of macro-F1. Other feature fusion including MFCC + DCT + VFDP; MFDP + DCT + MFCC did not performed better than the MFCC + DCT fusion, indicating that cepstral and modulation features together provide complementary information for optimal classification. In summary, the MLP findings reinforce the dominant contribution of spectral structure to language separability and demonstrate that non-linear modelling further amplifies discriminative patterns identified in earlier distributional analyses. Table 4 presents non-linear modelling consistently enhances performance for MFCC-dominant configurations, with MFCC + DCT achieving the highest overall classification accuracy and Macro-F1.

**Table 4.** Comprehensive Comparison of Linear SVM and Shallow MLP Performance Across All Acoustic Feature Configurations

| Feature Set | SVM | | MLP | |
|---|---|---|---|---|
| | Accuracy (%) | Macro-F1 (%) | Accuracy (%) | Macro-F1 (%) |
| LF | 84.72 ± 7.69 | 62.02 ± 15.21 | 84.15 ± 4.45 | 61.88 ± 16.00 |
| DCT | 85.54 ± 6.83 | 63.12 ± 15.30 | 84.22 ± 7.03 | 62.45 ± 16.73 |
| MFCC | 90.38 ± 3.70 | 87.28 ± 4.38 | 91.69 ± 3.97 | 87.66 ± 6.25 |
| MFCC+ DCT | 90.90 ± 4.47 | 88.32 ± 4.49 | **93.96 ± 4.16** | **92.42 ± 4.12** |
| LF+ DCT | 85.77 ± 7.05 | 62.79 ± 16.02 | 85.17 ± 5.02 | 62.56 ± 17.08 |
| LF+ MFCC | 88.70 ± 7.41 | 75.37 ± 15.41 | 92.06 ± 6.56 | 79.03 ± 16.30 |
| MFCC+ DCT +VFDP | 88.18 ± 7.80 | 75.45 ± 16.01 | 92.86 ± 7.09 | 81.30 ± 17.57 |
| MFDP+DCT+MFCC | 89.53 ± 8.01 | 76.17 ± 15.49 | 91.86 ± 8.24 | 80.01 ± 17.35 |

### 4.5 Discussion

The findings from this study indicate that Nyishi and Adi display quantifiable but structured acoustic differentiation, with systematic patterns emerging across both rhythmic and spectral domains. Threshold-based analyses showed consistent divergence in modulation frequency descriptors MFDP and VFDP with increased stability at higher thresholds and reduced sensitivity of peak count measures under conservative thresholding. Mixed-effects modelling also supported the importance of rhythmic frequency dispersion VFDP in differentiating the two languages. Although the degree of rhythmic differentiation was moderate, the consistency of these results across different analyses suggests that low-frequency modulation structure encodes systematic intra-branch variation, rather than random noise. This moderate rhythm differentiation is significant when considered in relation to the shared genealogy and prosodic structure of the languages within the subfamily. As closely related languages, they exhibit similar macro-level temporal organization, which constrains the extent of rhythm differentiation seen in the lower frequency amplitude modulation patterns. The similarity of rhythmic patterns, particularly in measures such as NDP at lower thresholds, does not indicate an absence of discrimination. In this context, frequency-based features provide more robust discrimination of intra-branch variation.

Spectral representations, on the other hand, exhibit greater levels of differentiation. The presence of a significant language effect in MFCC-PC1, along with the substantially higher Bhattacharyya distance in the MFCC domain, indicates that global spectral patterns offer greater degrees of acoustic differentiation. These patterns are likely reflecting differences in segmental realisation and articulatory configurations, which are more variable across closely related languages than macro-temporal rhythmic structure. DCT-based modulation features occupy an intermediate position, with higher-order features showing selective sensitivity. Their combination with MFCC features improving classification accuracy, suggesting that modulation-domain information provides complementary structural information beyond standard cepstral features. The improvement observed in feature fusion can be understood in terms of the complementary nature of the acoustic representations, where MFCC features capture fine-grained spectral-phonological variation, while DCT based modulation features encode broader temporal-spectral structures.

The results of supervised classification tasks also reflect this hierarchical structure. While rhythm-only features yielded moderate accuracy of 84-85%, their consistent performance across models indicates that low-frequency modulation captures systematic, though constrained, variation. The higher performance of MFCC-dominant models reflects the greater variability of spectral-phonological features rather than a limitation of rhythmic modelling. Feature fusion results demonstrate that modulation-based and spectral representations capture complementary aspects of linguistic organisation, with combined features achieving the highest classification performance. The enhanced performance in feature set combinations like MFCC + DCT thus owes more to the complementarity between these representations rather than simple additive effect. The comparison between individual and combination feature sets

Table 4 can also be viewed as an implicit ablation analysis showing that while MFCC features give excellent performance by themselves, there is always enhancement in accuracy through addition of modulation-based features. The higher variation at the speaker level found in the mixed effects model is an outcome of normal inter-speaker variation in the acoustic realisation. Nevertheless, the adoption of speaker independent GroupKFold cross validation guarantees that speaker related features do not influence classification accuracy, as there is no speaker that occurs both in the training and testing datasets. Overall, the results from statistical modelling, distributional analysis, and classification experiments converge to support a hierarchical organisation of acoustic differentiation, in which spectral features encode fine-grained phonological variation, while low-frequency rhythmic features reflect more stable macro-temporal structure. These results thus underscore the importance of multidimensional acoustic modelling for closely related languages and establish the complementary utility of modulation-based features in the study of intra-branch differentiation.

## 5. Conclusion

The results show that there is a pattern of intra-branch variation, with low-frequency temporal modulation patterns showing a moderate but significant level of separation, while spectral patterns show greater acoustic divergence. The most robust feature set for discrimination was based on MFCCs, especially when complemented with DCT modulation descriptors, which indicates that the spectral-phonological pattern of divergence is more salient than macro-temporal patterning in this case. In general, the results of both statistical and machine learning analyses provide strong support for a hierarchical approach to modelling acoustic variation and the utility of multi-dimensional analysis in investigating subtle differentiation among closely related under-resourced Tibeto-Burman languages.

Future research can further probe the linguistic underpinnings of this hierarchical pattern of differentiation by incorporating more nuanced phonetic and prosodic analysis to better understand the relationship between spectral patterns and macro-temporal organisation. The extension of the current modulation-based framework to other Tani languages would allow for a more comprehensive acoustic modelling of similarity and difference within the subgroup. The incorporation of larger datasets and more sophisticated representation learning methods can also further improve intra-branch language modelling and help in the development of effective speech technologies for under-resourced languages of Northeast India.

**Data Availability:** Data will be made available on request.

**Declarations**

**Conflict of interest:** The authors declare that they have no competing interests.

**References**

1. Dauer, R. M. (1983). Stress-timing and syllable-timing reanalyzed. *Journal of phonetics*, *11*(1), 51-62.
2. Loukina, A., Kochanski, G., Rosner, B., Keane, E., & Shih, C. (2011). Rhythm measures and dimensions of durational variation in speech. *The Journal of the Acoustical Society of America*, *129*(5), 3258-3270.
3. Duarte, D., Galves, A., Lopes, N., & Maronna, R. (2001). The statistical analysis of acoustic correlates of speech rhythm. In *Workshop on Rhythmic patterns, parameter setting and language change, ZiF, University of Bielefeld. Can be downloaded from http://www. physik. unibielefeld. de/complexity/duarte. pdf*.
4. Abercrombie, D. (2019). *Elements of general phonetics*. Edinburgh University Press.
5. Wiget, L., White, L., Schuppler, B., Grenon, I., Rauch, O., & Mattys, S. L. (2010). How stable are acoustic metrics of contrastive speech rhythm? *The Journal of the Acoustical Society of America*, *127*(3), 1559-1569.
6. Leong, V., & Goswami, U. (2015). Acoustic-emergent phonology in the amplitude envelope of child-directed speech. *PloS one*, *10*(12), e0144411.
7. Tilsen, S., & Johnson, K. (2008). Low-frequency Fourier analysis of speech rhythm. *The Journal of the Acoustical Society of America*, *124*(2), EL34-EL39.
8. Greenberg, S., Carvey, H., Hitchcock, L., & Chang, S. (2003). Temporal properties of spontaneous speech—a syllable-centric perspective. *Journal of Phonetics*, *31*(3-4), 465-485.
9. Dafydd Gibbon and Peng Li. 2019. Quantifying and correlating rhythm formants in speech. In Proceedings of the International Symposium on Linguistic Patterns in Spontaneous Speech (LPSS'19). Taipei, Academia Sinica.
10. Ramus, F., Nespor, M., & Mehler, J. (1999). Correlates of linguistic rhythm in the speech signal. *Cognition, 73*(3), 265–292.
11. Grabe, E., & Low, E. L. (2002). Durational variability in speech and the rhythm class hypothesis. In *Papers in Laboratory Phonology 7* (pp. 515–546). Cambridge University Press.
12. White, L., & Mattys, S. L. (2007). Rhythmic typology and variation in first and second languages. *Amsterdam Studies in the Theory and History of Linguistic Science Series 4*, 237–257.
13. Nolan, F., & Asu, E. L. (2009). The pairwise variability index and coexisting rhythms in language. *Phonetica, 66*(1–2), 64–77.
14. Arvaniti, A. (2009). Rhythm, timing and the timing of rhythm. *Phonetica, 66*(1–2), 46–63.

15. Arvaniti, A. (2012). The usefulness of metrics in the quantification of speech rhythm. *Journal of Phonetics, 40*(3), 351–373.

16. Barbosa, P. A. (2002). Explaining cross-linguistic rhythmic variability via a coupled-oscillator model of rhythm production. In *Proceedings of Speech Prosody 2002* (pp. 163–166).

17. Lancia, L., Krasovitsky, G., & Stuntebeck, F. (2019). Coordinative patterns underlying cross-linguistic rhythmic differences. *Journal of Phonetics, 72*, 66–80.

18. Polyanskaya, L., Busà, M. G., & Ordin, M. (2020). Capturing cross-linguistic differences in macro-rhythm: The case of Italian and English. *Language and Speech, 63*(2), 242–263.

19. Tilsen, S., & Arvaniti, A. (2013). Speech rhythm analysis with decomposition of the amplitude envelope: Characterizing rhythmic patterns within and across languages. *The Journal of the Acoustical Society of America, 134*(1), 628–639.

20. Gibbon, D. (2019). Computational induction of prosodic structure. *arXiv preprint arXiv:1912.07050.*

21. Gibbon, D. (2020). Rhythm formants of story reading in standard Mandarin. *Chinese Journal of Phonetics, 14*, 1–16.

22. Gibbon, D. (2023). The rhythms of rhythm. *Journal of the International Phonetic Association, 53*(1), 233–265.

23. Gibbon, D., & Lin, X. (2019). Rhythm zone theory: Speech rhythms are physical after all. In *Approaches to the Study of Sound Structure and Speech* (pp. 109–128). Routledge.

24. Gogoi, P., Sarmah, P., & Prasanna, S. R. M. (2024). Cross-linguistic rhythm analysis of Mising and Assamese. *ACM Transactions on Asian and Low-Resource Language Information Processing, 23*(10), 1–18.

25. Gogoi, P., Kalita, S., Sarmah, P., & Prasanna, S. R. M. (2025). Exploring Rhythm Formant Analysis for Indic language classification. In *Proceedings of the 28th Conference of the Oriental COCOSDA International Committee for the Co-ordination and Standardisation of Speech Databases and Assessment Techniques (O-COCOSDA)* (pp. 1–6).

26. Chakraborty, J., Sinha, R., & Sarmah, P. (2024). Effects of rate of articulation in Rhythm Formant Analysis-based dialect classification. In *Proceedings of the International Conference on Asian Language Processing (IALP)* (pp. 417–422).

27. Sun, J. T. S. (2003). Tani languages. *The Sino-Tibetan Languages*, 456-66.

28. Post, M. W., & Sun, J. T. S. (2017). Tani languages. *The Sino-Tibetan languages*, 322-337.

29. Post, M. W. (2021). On reconstructing ethno-linguistic prehistory: The case of tani. *Crossing boundaries: Tibetan studies unlimited*, 311-40.

30. Dafydd Gibbon. 2019. The phonetic grounding of prosody: Analysis and visualisation tools. In Human Language Tech nology. Challenges for Computer Science and Linguistics: 9th Language and Technology Conference, LTC 2019, Poznan, Poland, May 17–19, 2019, Revised Selected Papers (Poznań, Poland), Zygmunt Vetulani, Patrick Paroubek, and Marek Kubis (Eds.). Springer-Verlag, Berlin, 28–45//doi.org/10.1007/978-3-031-05328-3_3.

31. Dafydd Gibbon. 2020. Computational induction of prosodic structure. Studies in Prosodic Grammar 6, 2 (2020), 1–46.
32. Gibbon, D. TIME, COHESION, STYLE: RHYTHM FORMANTS IN ORAL NARRATIVE.
33. Ahmed, N., Natarajan, T., & Rao, K. R. (1974). Discrete cosines transform. *IEEE transactions on Computers*, *100*(1), 90-93.
34. Rabiner, L., & Schafer, R. (2010). *Theory and applications of digital speech processing*. Prentice Hall Press.
35. Davis, S., & Mermelstein, P. (1980). Comparison of parametric representations for monosyllabic word recognition in continuously spoken sentences. *IEEE transactions on acoustics, speech, and signal processing*, *28*(4), 357-366.
36. Davis, S.B. and Mermelstein, P. (1980), "Comparison of parametric representation for monosyllabic word recognition in continuously spoken sentences," IEEE Trans. on ASSP, Aug. 1980.
37. Douglas Bates, Martin Mächler, Ben Bolker, and Steve Walker. 2015. Fitting linear mixed-effects models using lme4. Journal of Statistical Software 67, 1 (2015), 1–48.
38. Welch, B. L. (1947). The generalization of 'STUDENT'S'problem when several different population varlances are involved. *Biometrika*, *34*(1-2), 28-35.
39. Brereton, R. G., & Lloyd, G. R. (2010). Support vector machines for classification and regression. *Analyst*, *135*(2), 230-267.
40. Bourlard, H., & Wellekens, C. J. (1989). Speech pattern discrimination and multilayer perceptrons. *Computer Speech & Language*, *3*(1), 1-19.